\documentclass[11pt]{article}
\usepackage{graphicx}
\usepackage{amsmath}
\usepackage{amssymb}
\usepackage{amsxtra}

\title{Formation and Evolution of Antimatter Objects}
\author{Sattvik Yadav\\National Institute of Science Education and Research, India\\sattvik.yadav@niser.ac.in}
\date{November 1, 2025}

\begin{document}
\maketitle

\begin{abstract}
	The fundamental question of baryogenesis and the problem of matter-antimatter asymmetry motivate this study into the formation and evolution of antimatter objects in the early Universe. Hypothesize is the existence of isolated antimatter domains in a baryon-asymmetric Universe that survive until the era of first star formation ($Z \approx 20$). By assuming CPT-symmetry, the thermodynamics, mechanics, and energy dynamics of an antimatter gas cloud (composed of antihydrogen and antihelium) are treated symmetrically to their primordial matter counterparts. Analysis demonstrates the physical feasibility of the gravitational collapse process for a conservatively estimated antimatter domain ($\approx 5 \times 10^3 M_{\odot}$). The initial conditions easily satisfy the Jeans and Bonnor-Ebert mass criteria, indicating a high propensity for instability and runaway collapse. The subsequent dynamical evolution, driven by $\bar{H}_2$ cooling, is predicted to proceed identically to that of Population III star formation, leading to the formation of a dense, adiabatic anti-protostellar core. The theoretical viability of a true antistar hinges upon a critical assumption: the physical possibility of antinuclear fusion (e.g., the antiproton cycle) under extreme core conditions. Assuming this symmetry holds, the collapse is predicted to yield massive antistars ($\gtrsim 22 M_{\odot}$). This suggests that if antimatter domains formed in the early Universe, they likely underwent stellar formation. Observational constraints on the existence of these objects must rely on the detection of characteristic high-energy $\gamma$-ray or X-ray signals resulting from matter-antimatter annihilation at the domain boundaries or during mass accretion.
\end{abstract}

\section{\label{sec:intro}Introduction}

The problem of baryon-asymmetry in the present universe is a prominent field of study today. Theories developing upon the universe having equal amount of matter and antimatter from the in early stages of universe fail to reach the present observed number density of matter in the universe. Modern cosmology now has shown that there indeed exists a baryon asymmetry in the universe. \\
\\
The study presented in this report finds its basis by considering existence of regions of antimatter in a baryon-asymmetric universe called the antimatter domains. These are proposed to form in early stages of universe (at age of universe $\leq 10^{-6}$) through a phase transition which separate out matter and antimatter. Key features of these regions is the annihilation happening only at the boundary of these region. Thus as these regions evolve in time, particles at the boundary interact and annihilate. These are proposed to be of sizes up to $\sim 10^{12} M_{\odot}$ taking into account that there is continuous coalescing of different antimatter domains and these domains survive till the period of hydrogen recombination (or antihydrogen in this case)\cite{Chechetkin}. \\
\\
Thus if existence of such regions are possible, whether there is possibility of structure formation inside these regions is probed in this study. It covers the gas clouds and move to viability of star formation. Primary focus is upon such structures in the era of early universe at an age $\sim 10^{13} sec$ (Redshift $Z \approx 20$), because this period has pressure and temperature considered viable for formation of first stars and possibly antistars.  Note that the thermodynamics of antimatter is considered symmetric to the matter. This means that for a macroscopic region of antimatter, the laws of thermodynamics hold for them. In other words Equations \ref{eqn:thermodynamic_1} and \ref{eqn:thermodynamic_2} hold for antimatter regions.  

\begin{equation}
	\label{eqn:thermodynamic_1}
	dU = \delta Q - \delta W
\end{equation}

\begin{equation}
	\label{eqn:thermodynamic_2}
	S = \frac{Q_{rev}}{T}
\end{equation}

\noindent Antimatter is currently considered to be CP invariant though theories also exist classifying the matter-antimatter thermodynamics and giving us the temperature measure for antimatter by a observer in matter universe to be negative. This would affect the second law of thermodynamics (Equation \ref{eqn:thermodynamic_2}), though the first law (Equation \ref{eqn:thermodynamic_1}) would remain unaffected as it concerns itself with energy of the system which is positive for both matter and antimatter alike. Proceeding with the CP-invariant thermodynamics, the sections in the report are structured by first a physical description in terms of matter and the viability of that physics in the antimatter. 

\section{\label{sec:initial}Initial Conditions for Star Formation}
The process of star formation greatly depends upon initial conditions the cloud of matter has from which the star is born. For analysis of antimatter, instead of considering the amount of antimatter possible in the domains as mentioned before, pessimistically consider this region to have a total mass of $\approx 10^5 M_\odot$. This include the mass of antimatter and dark matter. Assuming $\Lambda$-CDM universe, the matter to dark matter ratio in these domains (or gas clouds) is extend this to antimatter as well. This implies that for such antimatter domains present in a $\Lambda$-CDM universe, the percentage by mass of the antimatter present is $5\%$ and the rest is filled by the dark matter. Thus, net antimatter available for the star formation is $\approx 5 \times 10^3 M_\odot$. The antimatter domain is assumed to have no matter domains surrounding it, thus no interaction occurs at the boundaries. \\
\\
Now consider the evolution of the matter counterpart of antimatter. In the early Universe at $Z \approx 20$, condensed regions of matter have already formed with an average number density of domains of $\sim 10^{5} cm^{-3}$ and a temperature of $1000K$ \cite{Yoshida 2006}. These condensed regions are stabilized by two primary opposing forces: the gravitational force acting inwards and the pressure exerted by the particles acting outwards. These regions at the start primarily contained species such as $e^-$, $H$, $H^{-}$, $H^{+}$, $He$, $He^{+}$, $He^{++}$, $H_2$, and $H_2^+$. Other molecular species are present but in trace quantities.\\
\\
A gas under hydrostatic equilibrium requires it to follow the condition in Equation \ref{eqn:PGHydrostatic_Equilibrium}. These domains have the density low enough so that the gas present in the domain can be approximated to be ideal gas. These domains thus will also be referred as gas clouds of matter or (antimatter when specified). Assume spherical symmetry of the gas cloud which at start is isothermal. $P$ is the local pressure, $\rho$ is the local density, $\phi$ is the gravitational potential energy and $r$ is the radial distance from center of gas cloud. This combined with Poisson's equation (Equation \ref{eqn:Poission}) gives a description of the relation between different mechanical and thermodynamic quantities for a gas cloud in hydrostatic equilibrium.

\begin{equation}
	\label{eqn:PGHydrostatic_Equilibrium}
	\frac{dP}{dr} = -\frac{d\phi}{dr}\rho
\end{equation}

\begin{equation}
	\label{eqn:Poission}
	\frac{1}{r^2}\frac{d}{dr}\left(r^2 \frac{d\phi}{dr}\right) = 4\pi G\rho
\end{equation}

\noindent To fully solve this system to find $P$, $\phi$ and $\rho$ require another equation. This a model function which describes the relation between $P$ and $\rho$.

\begin{equation}
	\label{eqn:polytropic}
	P = K\rho^\gamma \equiv K\rho^{1 + \frac{1}{n}}
\end{equation} 

\noindent This is called the polytropic relation. $K$, $\gamma$ and $n$ are constants. Note that for properly chosen values of $\gamma$ and $K$, it gives us the ideal gas equation. Assuming the gaseous constituents form a mixture of ideal gas and region encompassed is very large, this gives us the choice of $n\rightarrow \infty$. Thus solving equations \ref{eqn:PGHydrostatic_Equilibrium}, \ref{eqn:Poission} and \ref{eqn:polytropic}, Lane-Emden Equation is obtained (\ref{eqn:LaneEmbden}).

\begin{equation}
	\label{eqn:LaneEmbden}
	\frac{d^2w}{dz^2} + \frac{2}{z}\frac{dw}{dz} = e^{-w}\\
\end{equation}

\noindent Where,

\begin{eqnarray}
	z = Ar \:\:,\:\:
	A^2 = \frac{4\pi G \rho_c}{K} \:\:,\:\:
	\phi = Kw
\end{eqnarray}

\begin{figure}
	\centering
	\includegraphics[width=0.7\linewidth]{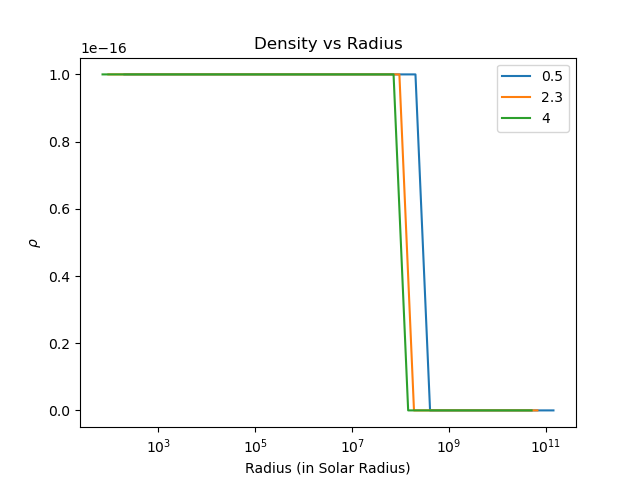}
	\caption{The figure show the variation of density of a cloud with ideal gas (for ideal gas, $n \rightarrow \infty$). The gas in the primordial clouds can be approximately considered as an ideal gas, since a low density gas occupy a large volume. The figure is obtained through plotting the Lane-Emden equation. Here diffenently coloured lines corresponds to different value of the mean molecular mass $\mu$. Blue line with $\mu = 0.5$ models the early universe, orange line with $\mu = 2.3$ models an average cloud in current universe and green line with $\mu = 4$ shows clouds with high metallicity. For a stable gas cloud with same boundary ccondition, it is see that the clouds were significatly bigger in the early universe.}
	\label{fig:mu-value-trend}
\end{figure}

\noindent $\rho_c$ is the central density. This differential equation is solved using the boundary condition by setting the central potential and central force 0. This equation was numerically solved, the results to which are shown in the Figure \ref{fig:mu-value-trend}.

\noindent At a macroscopic level, the thermodynamic quantity most relevant to us is the mean molecular mass ($\mu$) of the particles in gas cloud which is given by Equation \ref{eqn:mu}. This applies for the period of universe considered as the temperatures were high enough for most of the atoms to be ionized. $Z_i$ are the atomic number of the species, which corresponds to the free electrons for a neutral but ionised gas cloud and the $\mu_i$ is its molecular mass. For present universe, where in such gas clouds neutral atoms are present thereby $Z_i$ factor is absent. 

\begin{equation}
	\label{eqn:mu}
	\mu = \left( \sum_i \frac{X_i (1 + Z_i)}{\mu_i} \right)^{-1}
\end{equation} 

\section{\label{sec:process}Process of Star Formation}
\subsection{\label{subsec:onset}Onset of Collapse}
The initialization of collapse process happens when there are perturbation in the gas cloud. This perturbation may or may not lead to collapse depending upon the nature of perturbation or equivalently, the mass of matter upon which this perturbation is acting on. These perturbations in the gas cloud travel as sound wave, and have a finite time of propagation. In our current universe, these are produced by the shock waves emitted in supernovas and other highly energetic events. But in the early universe, such disturbances can be produced due to high energy particles or annihilation events, thus paving way for star formation. In terms of mass, the minimum mass of gas required such that a perturbation of sufficient energy is enough to start a run collapse (keeping the assumptions used before) is given by the Jeans' condition

\begin{equation}
	\label{eqn:Jeans}
	M_J = 1.1 M_{\odot} (\frac{T}{10K})^{\frac{3}{2}} \left(\frac{\rho}{10^{-19}g cm^{-3}}\right)^{-\frac{1}{2}} \left(\frac{\mu}{2.3}\right)^{-\frac{3}{2}}
\end{equation}

\noindent A more detailed calculation through a better boundary condition encompassing the finite gas cloud gives us the Bonnor-Ebert condition \cite{Larson 1969}

\begin{equation}
	\label{eqn:BE2}
	M_{BE} =  1.18 \frac{R_H^2}{\mu^2 G^{\frac{3}{2}}} T^2 (P^*)^{-\frac{1}{2}} M_{\odot}
\end{equation}

\noindent For gas with lesser mass than this limit, the perturbation form a stable wave and propagate without causing a collapse. Upon calculation using the initial conditions mentioned in the previous section, the following mass values is obtained.

\begin{table}[h]
	\centering
	\label{tab:mass calculation}
	\resizebox{0.8\linewidth}{!}{%
		\begin{tabular}{|c|c|c|}
			\hline
			\textbf{$\mu$} & \textbf{\begin{tabular}[c]{@{}c@{}}Jeans Mass\\ (in $M_{\odot}$)\end{tabular}} & \textbf{\begin{tabular}[c]{@{}c@{}}Bonnor-Ebert Mass\\ (in $M_{\odot}$)\end{tabular}} \\ \hline
			0.5 & 5581.491579 & 5689.938953 \\ \hline
			2.3 & 565.735623 & 1236.943251 \\ \hline
			4.0 & 246.669409 & 711.242369 \\ \hline
		\end{tabular}%
	}
		\caption{Calculated values for the Jeans Mass (JM) and Bonnor Ebert Mass (BEM) for different values of the mean molecular mass $\mu$. Note that the collapse mass required is approximately equal to the JM and BEM of the initial gas cloud of antimatter taken. The $\mu = 0.5$ value is the mean molecular mass calculated for antimatter based upon the chemical abundance of the species mentioned earlier and by considering atomic hydrogen in majority.}
\end{table}

\noindent Once the process of collapse starts, the gas cloud radiates energy. Due to extremely low opacity gas in the cloud in the early universe, as it primarily consisted of species of hydrogen and helium, the rate of cooling is very high. This cools the gas cloud from an initial temperature of $1000K$ till $200K$. The energy radiated is primarily provided by highly exothermic reaction forming $H_2$ from $H^-$. It goes as follows:

\begin{eqnarray*}
	H + e^- \rightarrow H^- + h\nu \\
	H^- + H \rightarrow H_2 + e
\end{eqnarray*}

\noindent This reaction also becomes the primary source of formation of large amount of $H_2$ in the early universe and for the sharp reduction of the temperature of clouds. 

\subsection{\label{subsec:sim} Collapse of Gas Cloud}
Once the collapse start, the system is no longer in an equilibrium. Note that even though cooling of gas is occuring along its collapse, the process can still be considered quasi-isothermal. This is so because the net time required for the gas to free fall to the central point is $10^7$ years, which is even greater considering the pressure acting from the gas. On the other hand, the time taken for the gas to reach thermal equilibrium is $10$ years which is much less that it. Thus the isothermal assumption is safe to assume for long timescales of collapse. From the conservation of mass in a volume element in the cloud, continuity equation is used, which goes as follows:

\begin{equation}
	\frac{\partial m}{\partial t} + 4\pi r^2 v \rho = 0
\end{equation}

\noindent Second, from the Newton's Laws of motion for element of mass $m$. Since the equilibrium is broken, the forces acting will be unbalanced and thus a net acceleration given by
\begin{equation}
	m \frac{d\vec{v}}{dt} = \vec{F}_{Gravitational} + \vec{F}_{Pressure}
\end{equation}

\noindent Third, conservation of angular momentum, 
\begin{equation}
	\vec{N} = \sum_{\text{over all elements}} \vec{r}_i \times (\vec{F}_{Gravitational} + \vec{F}_{Pressure})
\end{equation}

\noindent Fourth, is the local change in the energy,
\begin{equation}
	\frac{du}{dt} + P \frac{d}{dt} \left(\frac{1}{\rho}\right) + \frac{1}{4 \pi \rho r^2}\frac{\partial\Lambda}{\partial r}
\end{equation}

\noindent Here $\Lambda$ is the rate of cooling of the gas. This quantity is very hard to construct analytically as the it depend upon the opacity, density and chemical nature of various species present in the cloud. Some theoretical calculation and numerical fitting corrections over experimental data obtained in the laboratory gives us the estimate of this cooling as follows \cite{Hollenbach 1979}

\begin{equation}
		\Lambda_{LTE} = \Lambda_{rot} + \Lambda_{vib}
\end{equation}

\begin{equation}
	\Lambda_{rot} = \frac{9.5 \times 10^{-22} (T_3)^{3.76}}{1 + 0.12 (T_3)^{2.1}} e^{-\left(\frac{0.13}{T_3}\right)^3} + 3 \times 10^{-24} e^{-\left(\frac{0.51}{T_3}\right)} 
\end{equation}

\begin{equation}
	\Lambda_{rot} = 6.7 \times 10^{-19} e^{-\left(\frac{5.86}{T_3}\right)} + 1.6 \times 10^{-18} e^{-\left(\frac{11.7}{T_3}\right)} 
\end{equation}

\noindent Here $T_3 = \frac{T}{1000 K}$. $\Lambda_{LTE}$ is the Local Thermodynamic Cooling which is due to the de-excitation of different species from a higher energy state to a lower state. Along with this, contribution by Emission due to collision of atoms are present. It depends upon the temperature and pressure of gas. Approximate numerically modeled function is given as follows \cite{Omukai 2001}

\begin{equation}
	\Lambda_{CIE} = 10^{-116.6 + 96.34 log T - 47.153 (log T)^2 +10.744 (log T)^3 - 0.916 (log T)^4}
\end{equation}

\noindent Using these functions, the process of a gas cloud made up of matter has been simulated by studies \cite{Yoshida 2006}.

\begin{figure}[h]
	\centering
	\includegraphics[width=\linewidth]{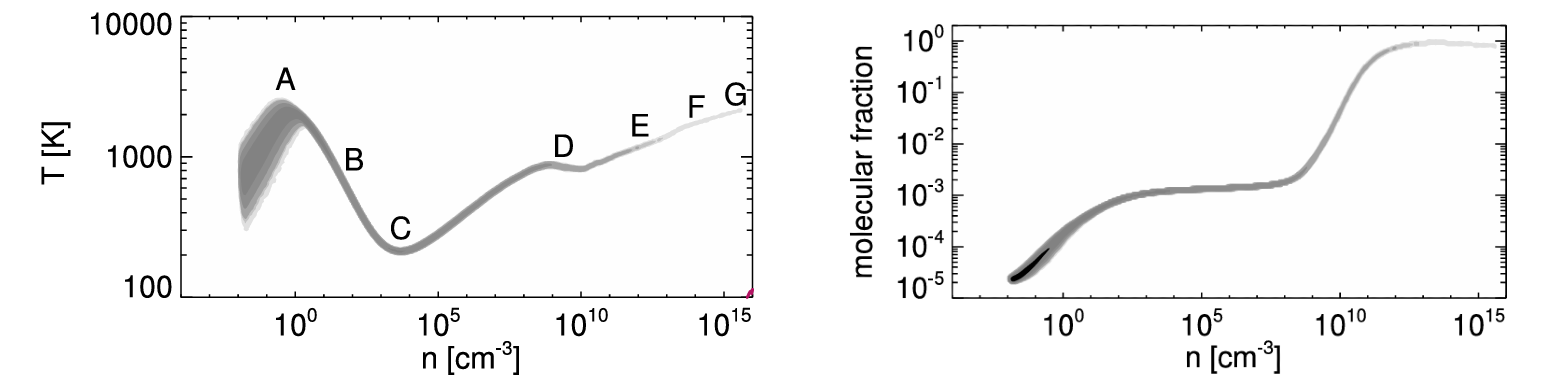}
	\caption{Variation of the temperature and molecular fraction with the increase in number density as simulated for primordial matter stars (Yoshida et al., 2006).}
	\label{fig:image}
\end{figure}

\noindent As the process of collapse goes on, when the number density of about $10^9 cm^{-3}$ is reached, the rate of $H_2$ formation is further enhanced. This is due conditions being right for the 3-body reaction of H to start.

\begin{eqnarray*}
	H + H + H \rightarrow H_2 + H \\
\end{eqnarray*}

\begin{eqnarray*}
	H_2 + H \rightarrow 2H_2
\end{eqnarray*}

\noindent While at this stage in the current Universe, the gas is opaque enough to trap the heat, thereby reducing the rate of cooling, since the clouds in the early Universe still lacked sufficient opacity, the cooling and the rate of collapse were still very high. When the number density reaches $10^{12} cm^{-3}$, the gas finally starts becoming opaque to the outgoing radiation and thus the temperature of the gas cloud also starts increasing as seen in Figure \ref{fig:image}. The inner regions of this cloud are shielded the most by increased opacity of the outer layers of the collapsing cloud. These regions later form the core of the star, now follows an adiabatic process instead of an quasi-isothermal one. The core temperature also becomes high enough to ionise the molecular species to the atomic ones. 

\subsection{\label{subsec:proto}Protostar Formation and Main Sequence}
\begin{figure}[h]
	\centering
	\includegraphics[width=\linewidth]{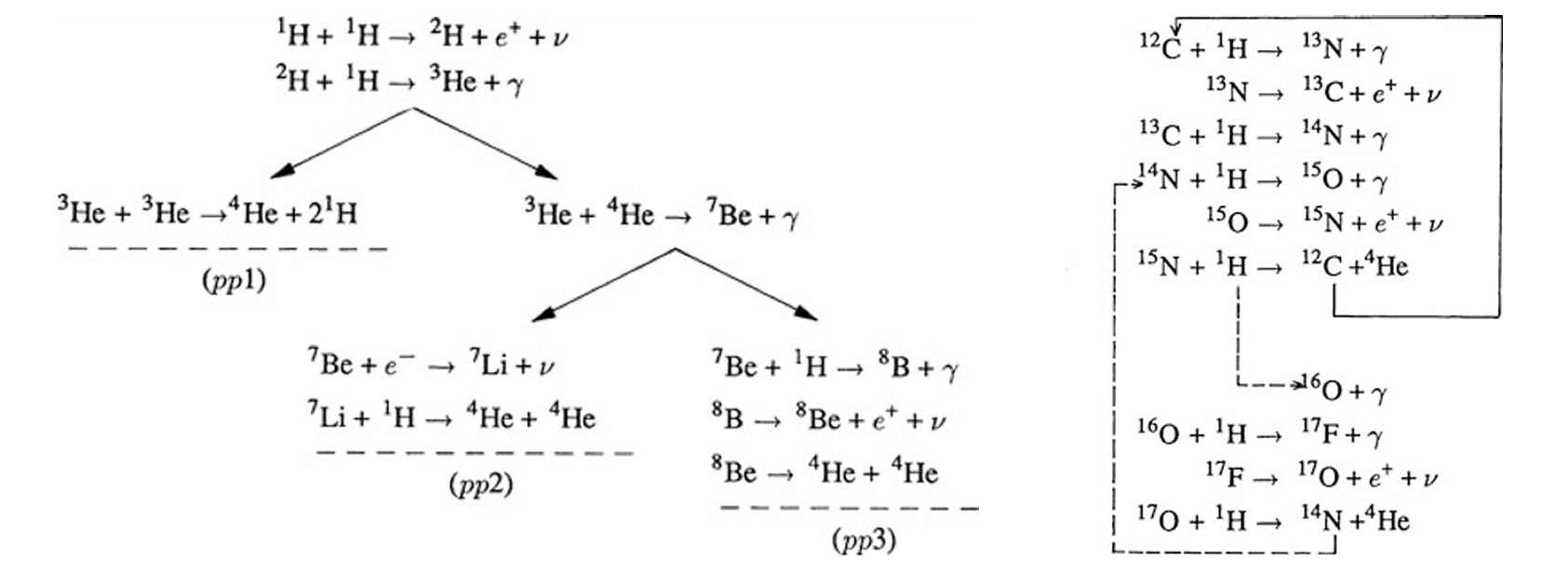}
	\caption{The following figure shows the major nuclear reactions that start to occur in the core during the nuclear burning. The set of reactions on the left is the ones that start first. The set of reactions on the right is the one that becomes dominant after an increase in metallicity due to fusion reactions (Kippenhahn \& Weigert, 2013).}
	\label{fig:stellar-reactions}
\end{figure}

Once this reionisation of molecular species occurs, in the adiabatic conditions of the core has the tendency to have nuclear fusion. Presence of two cycles are key for the nuclear burning which are shown in the Figure \ref{fig:stellar-reactions}.

In this stage of cycle, nuclear burning is starting in certain areas of the core while in-falling of matter still occurs. This is called the protostellar stage and the adiabatic core with radiating surrounding gas is enveloped by a region (or more precisely a disc) of matter. Now for this protostar to develop in an actual star, the temperature and pressure should be right. The phenomenon of a protostar core becoming degenerate and failing to ignite is the defining characteristic of brown dwarf formation. During the early stages of evolution, a protostar contracts and must release gravitational energy, leading to heating of its central core, following the typical behavior of an ideal gas sphere. If the central material consists of an ideal gas, further contraction leads to higher temperatures. the equation governing this is

\begin{equation}
	\label{eqn:central temp p}
	\frac{dT_c}{T_c} = \frac{4\alpha - 3}{3\delta}\frac{d\rho_c}{\rho_c}
\end{equation}

\noindent For the core which remains ideal and thus transition to have nuclear fusion have $\alpha = \delta = 1$. However, if the protostar's mass is too low, the central density increases rapidly enough that the electron gas becomes degenerate before the core can reach the necessary temperature for stable hydrogen burning (approximately $10^7$ K). When degeneracy dominates the equation of state, the pressure support becomes effectively independent of the core temperature. This transition to degenerate matter removes the core's ability to self-regulate. Without this mechanism, the contraction ceases to cause the necessary continuous heating, effectively halting the evolutionary track at a maximum temperature that is too low for fusion. The object, unable to achieve the thermal equilibrium characteristic of true stars, is supported purely by the non-thermal pressure of the degenerate electrons and begins to cool, settling as a brown dwarf. Based upon this density and temperature, the calculation done gives us the minimum mass of the protostar to be $0.08 M_{\odot}$. 

\begin{figure}[h]
	\centering
	\includegraphics[width=0.7\linewidth]{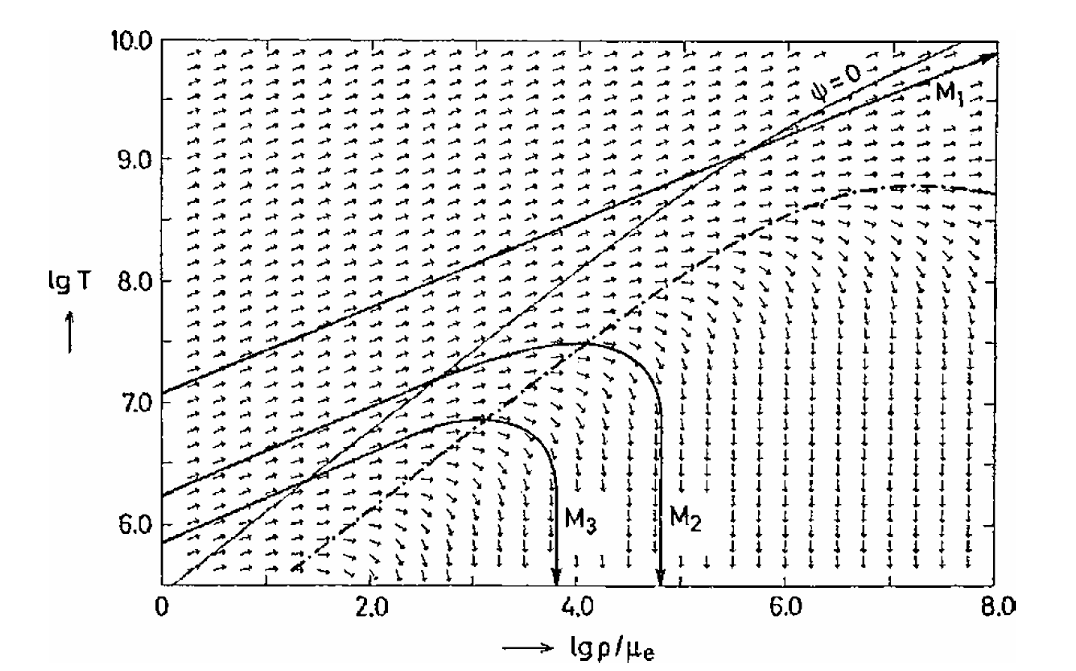}
	\caption{Graph showing the region determining whether the core becomes degenerate or not. The graph plots the Equation \ref{eqn:central temp p}. There is existence of two regions. One in the bottom right is due to the creation of electron degeneracy and thus represents a region where if the curve of stellar evolution enters, the star forms a brown dwarf (Kippenhahn \& Weigert, 2013).}
	\label{fig:lntc-vs-ln-pc-ploy}
\end{figure}

\noindent Note that for stars in early universe, due to high infalling matter the mass of the protostars results from the initial conditions to be approximately $1 M_{\odot}$ which is larger than what is required for protostar to go to main sequence \cite{Yoshida 2012}. Once the star reaches this stage high amount of nuclear burning starts which creates a strong radiation feedback. This stops and blows away further infall of the matter. Various studies using similar initial condition but with much more refined physics have given approximation or bounds over the value of the star formed after the end of star formation process in the early universe. Yoshida et. al. \cite{Yoshida 2006} gives $\approx 100 M_{\odot}$, Yoshida et. al. \cite{Yoshida 2012} gives $\approx 42 M_{\odot}$ and Stacy et. al \cite{Stacy 2012} gives a lower bound $\geq 22 M_{\odot}$.

\section{\label{sec:anti}Case of Antimatter}
Shifting the focus back to antimatter. Due to the symmetric nature with matter in terms of thermodynamics, mechanics and energy dynamics, since all the processes before the protostar formation and the main sequence depends on these quantities, for domains of antimatter, such process is feasible. Note that the possibility of atomic antihydrogen reaction similar to that of hydrogen is assumed here. The major hurdle occurs when the start of nuclear burning in the protostar is considered. For its antimatter counterpart, existence of such reactions is still under study. Thus if such reactions are possible, even though the initial conditions were pessimistic values, such antistars may be possible. Note that this study considers a lot of assumptions, relaxing those is still under study. Nevertheless initial values are promising for search of such stars in the universe.

\section{\label{sec:feasibiltiy}Feasibility}
\subsection{\label{subsec:detect}Detection of Antistars}
For detection of such structure in the universe, stages and properties of antistars can again be considered. The existence of possible antistellar objects depends upon their detection from the data obtained. Search for matter-antimatter annihilation products include high energy gamma ray radiation or possible transitions of certain species such as $p\bar{p}$, $p\bar{He}$ or $He\bar{He}$ in X-Ray regime \cite{Blinnikov 2015} \cite{Dupourqué 2021}.

\subsection{\label{subsec:bigstars}Possibility of Such Stars}
Feasibility of such antistars also depend upon whether there is a possibility of such stars in matter regime. There is no direct observational data for these first stars as due to the age of the universe and their large size, they would have had a short lifespan. At the same time, second generation of stars have been observed with very low metallicity and very high mass, hinting at the possibility of such massive stars .

\section{\label{sec:conclusion}Conclusion}
The study systematically explored the viability of antistar formation within hypothesized antimatter domains in a baryon-asymmetric universe. By assuming CP-invariant thermodynamics, the physics of antihydrogen and antihelium gas clouds was treated symmetrically to their well-studied matter counterparts.\\
\\
\noindent The analysis shows that the initial conditions for star formation in the early universe, specifically at $Z \approx 20$, are highly conducive to the collapse of antimatter gas clouds. Using the properties of primordial matter clouds ($\mu \approx 0.5$, $T \approx 1000 \text{ K}$, $\rho \sim 10^5 \text{ cm}^{-3}$), the calculated Jeans Mass ($M_J \approx 5581 M_{\odot}$) and Bonnor-Ebert Mass ($M_{BE} \approx 5690 M_{\odot}$) for $\mu=0.5$ are remarkably close to the conservatively estimated total antimatter mass available for collapse ($\approx 5 \times 10^3 M_{\odot}$). This suggests that if antimatter domains of the size and density described exist, they satisfy the mass requirement for gravitational instability and subsequent collapse.\\
\\
\noindent Furthermore, the processes driving collapse---cooling via $\bar{H}_2$ formation, the transition to three-body reactions, and the eventual onset of opacity at high densities---are governed by fundamental laws of thermodynamics and gravity that are assumed to hold true for antimatter. This symmetry strongly suggests that the dynamical evolution would proceed identically to that of primordial matter protostars, leading to the formation of massive antistars with predicted masses in the range of $\gtrsim 22 M_{\odot}$ \cite{Stacy 2012}. Crucially, these massive objects would successfully overcome the electron degeneracy pressure to ignite nuclear fusion.\\
\\
\noindent The entire collapse, from the onset of instability through the quasi-isothermal phase driven by $\bar{H}_2$ cooling to the formation of an opaque, adiabatic protostar core, is robustly feasible due to the symmetry in thermodynamics and mechanics. The formation of a stable, long-lived antistar, however, hinges critically on one key assumption: the feasibility of anti-nuclear fusion (e.g., anti-proton-anti-proton cycle) occurring under the high-temperature and high-pressure conditions of the anti-protostellar core.\\
\\
\noindent While the feasibility of detecting antistars remains highly challenging, future searches for high-energy gamma-ray and X-ray annihilation signatures, particularly at the boundaries of gas clouds and during phases of mass accretion onto antistars, represent the only immediate avenue for observational proof. The confirmation of such massive stellar-sized antimatter objects would not only validate models of early-universe phase transitions but would also provide a crucial empirical constraint on the symmetry of the laws governing nuclear energy generation.

\section*{Acknowledgments}
I would like to thank Prof. Dr. Maxim Yu. Khlopov and the entire organising committee for giving me an opportunity to present in the Bled Workshop 2025. This has been a wonderful opportunity for me to get more insight to this work as well as to know more about various fields of research by amazing speakers from across the world.


\end{document}